\documentclass[]{article}

\usepackage{graphicx}
\usepackage{amssymb}
\usepackage{html}

\newtheorem{theorem}{Theorem}

\newtheorem{definition}[theorem]{Definition}

\newtheorem{proposition}[theorem]{Proposition}
\newtheorem{remark}[theorem]{Remark}

\newenvironment{proof}[1][Proof]{\noindent\textbf{#1.} }{\ \rule{0.5em}{0.5em}}

\newcommand\overset{\stackrel}

\begin{document}

\title{Relativistic simultaneity and causality}

\author{V. J. Bol\'os$^1$, V. Liern$^2$, J. Olivert$^3$ \\ \\
{\small $^1$ Dpto. de Matem\'aticas, Facultad de Ciencias,
Universidad de Extremadura.}\\
{\small Avda. de Elvas s/n. 06071--Badajoz, Spain.}\\
{\small e-mail\textup{: \texttt{vjbolos@unex.es}}}
\\ \\
{\small $^2$ Departament d'Economia Financera i Matem\`atica,
Universitat de Val\`encia.}\\
{\small Avda. Tarongers s/n. 46071--Valencia, Spain.}\\
{\small e-mail\textup{: \texttt{vicente.liern@uv.es}}}
\\ \\
{\small $^3$ Departament d'Astronomia i Astrof\'{\i}sica, Universitat de Val\`encia.}\\
{\small C/ Dr. Moliner s/n. 46100--Burjassot (Valencia), Spain.}\\
{\small e-mail\textup{: \texttt{joaquin.olivert@uv.es}}}}

\date{March 9, 2005}

\maketitle

\begin{abstract}
We analyze two types of relativistic simultaneity associated to an
observer: the spacelike simultaneity, given by Landau
submanifolds, and the lightlike simultaneity (also known as
observed simultaneity), given by past-pointing horismos
submanifolds. We study some geometrical conditions to ensure that
Landau submanifolds are spacelike and we prove that horismos
submanifolds are always lightlike. Finally, we establish some
conditions to guarantee the existence of foliations in the
space-time whose leaves are these submanifolds of simultaneity
generated by an observer.
\end{abstract}

\section{Introduction}

It is well known that some problems related with simultaneity have
not been solved yet. A great number of works treat the local
character of relativistic simultaneity accepting that Landau
submanifolds \cite{Oliv80} generated by an observer are leaves of
a spacelike foliation. However, the fulfillment of this property
cannot be ensured on any neighborhood without assuming some
additional geometrical conditions. Therefore, when working on a
neighborhood where this property does not hold, some difficulties
in setting a successful dynamical study arise. The study of some
of these conditions is the main objective of this paper.

In this work, we consider two types of simultaneities:\
\textit{spacelike simultaneity}, which describes those events that
are simultaneous in the local inertial proper system of the
observer; and \textit{lightlike} (or \textit{observed})
\textit{simultaneity}, which describes those events which the
observer observes as simultaneous although they are not
simultaneous in its local inertial proper system. The sets of
spacelike simultaneous events and lightlike simultaneous events
determine the Landau submanifolds and the past-pointing horismos
\cite{Beem81} submanifolds respectively.

Our next concern is the causality related to these types of
simultaneity, because we should be able to guarantee, for
instance, that Landau submanifolds are spacelike in a given
neighborhood. For this, we introduce a new concept, the
\textit{tangential causality}, more general than causality, and we
prove that every Landau submanifold $L_{p,u}$ is $p$-tangentially
spacelike, but it is not necessarily spacelike. On the other hand,
we prove that every horismos submanifold $E_{p}$ is
$p$-tangentially lightlike and lightlike.

In physics, it is usual to work with synchronizable timelike
vector fields. In most cases, the leaves of the orthogonal
foliation are considered \textquotedblleft simultaneity
submanifolds\textquotedblright . In this work it is proved that
given an observer $\beta $ (i.e. a $C^{\infty }$ timelike curve),
there exists, on a small enough tubular neighborhood of this
observer, a synchronizable timelike vector field containing the
4-velocity of $\beta $. Moreover, this vector field is orthogonal
to a Landau submanifolds foliation. But we can not assure the
existence of these kind of vector fields on any convex normal
neighborhood. On the other hand, it is also proved that given an
observer, there exists a foliation whose leaves are past-pointing
horismos submanifolds of events of the observer, and it is well
defined on any convex normal neighborhood.

\section{Preliminary Concepts}

In what follows, $\left( \mathcal{M},g\right) $ will be a
4-dimensional lorentzian space-time manifold. Given $v$ a vector,
$v^{\bot }$ will denote the subspace orthogonal to $v$.

Let $p\in \mathcal{M}$. An open neighborhood $\mathcal{N}_{0}$ of
the origin in $T_{p}\mathcal{M}$ is said to be \textbf{normal} if
the following conditions hold:

\begin{description}
\item[(i)] the mapping $\exp _{p}:\mathcal{N}_{0}\rightarrow
\mathcal{N}_{p}$ is a diffeomorphism, where $\mathcal{N}_{p}$ is
an open neighborhood of $p$.

\item[(ii)] given $X\in \mathcal{N}_{0}$ and $t\in \left[
0,1\right] $ we have that $tX\in \mathcal{N}_{0}$.
\end{description}

For a given event $p\in \mathcal{M}$, an open neighborhood
$\mathcal{N}_{p}$
of $p$ is a \textbf{normal neighborhood of $p$} if $\mathcal{N}_{p}=\exp _{p}%
\mathcal{N}_{0}$, where $\mathcal{N}_{0}$ is a normal neighborhood
of the origin in $T_{p}\mathcal{M}$. Finally, an open set
$\mathcal{V}\neq \emptyset $ in $\mathcal{M}$, which is a normal
neighborhood of each one of its points, is a \textbf{convex normal
neighborhood}.

These neighborhoods are useful to obtain a dynamical study of
simultaneity. Moreover, the Whitehead Lemma asserts that given
$p\in \mathcal{M}$ and a neighborhood $\mathcal{U}$ of $p$, there
always exists a convex normal neighborhood $\mathcal{V}$ of $p$
such that $\mathcal{V}\subset \mathcal{U}$ \cite{Sach77}. Since we
are not going to make a global study, we will consider the
space-time $\mathcal{M}$ as a convex normal neighborhood in order
to simplify. So, given two events in $\mathcal{M}$, there exists a
unique geodesic containing them.

\begin{figure}[tbp]
\begin{center}
\includegraphics[width=0.75\textwidth]{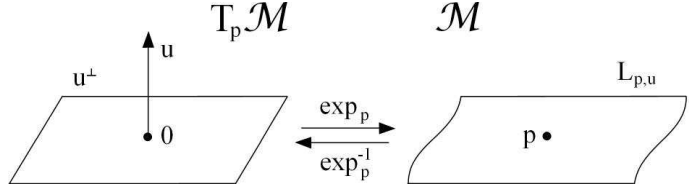}
\end{center}
\caption{Construction of the Landau submanifold $L_{p,u}$ by means
of the exponential map.} \label{figlandau}
\end{figure}

We are going to introduce two static ways to analyze simultaneity:
Landau and past-pointing horismos submanifolds. Given $u\in
T_{p}\mathcal{M}$ the 4-velocity of an observer at $p$, and the
metric tensor field $g$, we consider the submersion $\Phi
:\mathcal{M}\rightarrow
\mathbb{R}
$ given by $\Phi \left( q\right) :=g\left( \exp
_{p}^{-1}q,u\right) $. The
fiber%
\begin{equation}
L_{p,u}:=\Phi ^{-1}\left( 0\right)  \label{landau}
\end{equation}%
is a regular 3-dimensional submanifold called \textbf{Landau submanifold of }%
$\left( p,u\right) $. In other words,
\[
L_{p,u}=\exp _{p}u^{\bot }
\]
(see Figure \ref{figlandau}).

The next result is given in \cite{Oliv80}:

\begin{theorem}
Given $u\in T_{p}\mathcal{M}$ the 4-velocity of an observer at
$p$, there
exists a unique regular 3-dimensional submanifold $L_{p,u}$ such that $%
T_{p}L_{p,u}=u^{\bot }$ and whose points are simultaneous with $p$
in the local inertial proper system of $p$.
\end{theorem}

\begin{figure}[tbp]
\begin{center}
\includegraphics[width=0.65\textwidth]{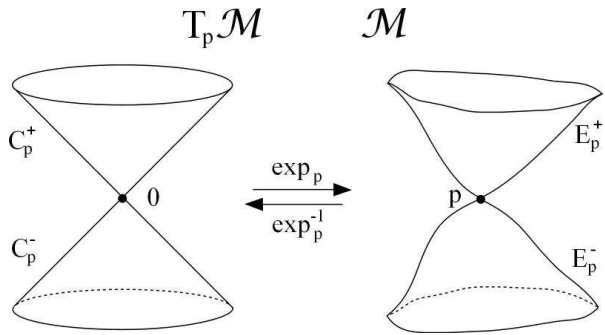}
\end{center}
\caption{Construction of the horismos submanifolds $E_p^-$ and
$E_p^+$ by means of the exponential map.} \label{fighorismos}
\end{figure}

On the other hand, defining the submersion $\varphi
:\mathcal{M}-\left\{ p\right\} \rightarrow
\mathbb{R}
$ given by $\varphi \left( q\right) :=g\left( \exp _{p}^{-1}q,\exp
_{p}^{-1}q\right) $, the fiber%
\begin{equation}
E_{p}:=\varphi ^{-1}\left( 0\right)  \label{horismos}
\end{equation}%
is a regular 3-dimensional submanifold, called \textbf{horismos
submanifold of }$p$, which has two connected components
\cite{Sach77}. We will call
\textbf{past-pointing} (respectively \textbf{future-pointing}) \textbf{%
horismos submanifold of }$p$, $E_{p}^{-}$ (resp. $E_{p}^{+}$), to
the connected component of (\ref{horismos}) in which, for each
event $q\in \mathcal{M}-\left\{ p\right\} $, the preimage $\exp
_{p}^{-1}q$ is a past-pointing (respectively future-pointing)
lightlike vector. In other
words,%
\[
E_{p}^{-}=\exp _{p}C_{p}^{-}~~~;~~~E_{p}^{+}=\exp _{p}C_{p}^{+},
\]
where $C_{p}^{-}$ and $C_{p}^{+}$ are the past-pointing and the
future-pointing light cones of $T_{p}\mathcal{M}$ respectively
(see Figure \ref{fighorismos}).

The events in a Landau submanifold $L_{p,u}$ are simultaneous with
$p$ in
the local inertial proper system of $p$ (i.e. they are synchronous with $p$%
), but they are not observed as \textit{simultaneous} (in the
sense that they are not observed \textit{at the same instant}) by
any observer. On the other hand, the events in a past-pointing
horismos submanifold $E_{p}^{-}$ are observed as simultaneous by
any observer at $p$, since they belong to
light rays which arrive at $p$. So, a Landau submanifold $L_{p,u}$ defines an%
\textit{\ intrinsic }simultaneity for an observer with 4-velocity $u$ at $p$%
, and a past-pointing horismos submanifold $E_{p}^{-}$ defines
an\textit{\ observed }simultaneity for any observer at $p$. In
general, we will call both, Landau and past-pointing horismos
submanifolds, \textbf{simultaneity submanifolds}.

\section{Tangential causality}

Our aim is to study the causality of simultaneity submanifolds,
specially in which cases a Landau submanifold is spacelike and in
which cases a horismos submanifold is lightlike. But first, we are
going to introduce a new concept called \textquotedblleft
tangential causality\textquotedblright , because some results on
tangential causality will be useful to study causality in Section
\ref{sec4}.

Let $V$ be a 4-dimensional vector space regarded as a $C^{\infty
}$ manifold. It can be canonically identified with any of its
tangent spaces: for each $v\in V$ there exists a unique
isomorphism $\phi
_{v}:T_{v}V\rightarrow V$ such that%
\begin{equation}
\omega \left( \phi _{v}w\right) =w\left( \omega \right)
\label{caniso}
\end{equation}%
for all $w\in T_{v}V$ and for all $\omega \in V^{\ast }$.

If, moreover, $\left( V,g\right) $ is a lorentzian vector space,
we define a
$\left( 0,2\right) $-tensor field $\mathbf{g}$ on $V$ by%
\begin{equation}
\mathbf{g}\left( w,w^{\prime }\right) :=g\left( \phi _{v}w,\phi
_{v}w^{\prime }\right)  \label{gbold}
\end{equation}%
where $v\in V$ and $w,w^{\prime }\in T_{v}V$. Then, $\left( V,\mathbf{g}%
\right) $ is a lorentzian manifold \cite{Sach77}.

Let $p$ be in $\mathcal{M}$, we can apply this to
$T_{p}\mathcal{M}$ because
$\left( T_{p}\mathcal{M},g_{p}\right) $ is a lorentzian vector space, where $%
g_{p}:=g|_{T_{p}\mathcal{M}}$. So, for each $v\in
T_{p}\mathcal{M}$ we can
define a canonical isomorphism $\phi _{v}$ of $T_{v}\left( T_{p}\mathcal{M}%
\right) $ onto $T_{p}\mathcal{M}$ satisfying (\ref{caniso}). If we define $%
\mathbf{g}_{p}$ on $T_{v}\left( T_{p}\mathcal{M}\right) $ from
$g_{p}$
(according to (\ref{gbold})), then $\left( T_{p}\mathcal{M},\mathbf{g}%
_{p}\right) $ is a lorentzian manifold.

\begin{definition}
Let $N$ be a regular submanifold of $\mathcal{M}$ and $p\in N$. The $p$%
\textbf{-tangential submanifold of }$N$ is%
\begin{equation}
\exp _{p}^{-1}N  \label{ptangsub}
\end{equation}%
considered as a regular submanifold in $T_{p}\mathcal{M}$.

Given $q\in N$, we define the $p$\textbf{-tangential causality of }$N$%
\textbf{\ at }$q$ as the causality of $\exp _{p}^{-1}N$ at $\exp _{p}^{-1}q$%
, using $\mathbf{g}_{p}$. If this causality is the same at every point of $%
\exp _{p}^{-1}N$, then we define the $p$\textbf{-tangential
causality of }$N$ as the causality of $\exp _{p}^{-1}N$ at any
point.
\end{definition}

The $p$-tangential causality can be interpreted as an
\textquotedblleft observed causality at $p$\textquotedblright ,
because an observer detects the events of the space-time through
its tangent space.

It is easy to prove that
\begin{equation}
T_{q}N=\exp _{p\ast v}\left( T_{v}\left( \exp _{p}^{-1}N\right)
\right) , \label{expest}
\end{equation}%
where $N$ is a regular submanifold of $\mathcal{M}$, $p,q\in N$,
and $v=\exp
_{p}^{-1}q$. As a particular case, since $\exp _{p\ast 0}=\phi _{0}$ \cite%
{Beem81}, we have%
\[
T_{p}N=\phi _{0}\left( T_{0}\left( \exp _{p}^{-1}N\right) \right)
.
\]
Then, the causality of $N$ at $p$ coincides with the
$p$-tangential
causality of $N$ at $0$. However, given $v\in \exp _{p}^{-1}N$ such that $%
v\neq 0$, the causality of $N$ at $p$ is not necessarily the same as the $p$%
-tangential causality of $N$ at $v$.

Applying the tangential causality concept to simultaneity
submanifolds, we obtain the next result:

\begin{proposition}
\label{proptangcau} Given $p\in \mathcal{M}$, and $u\in
T_{p}\mathcal{M}$ the 4-velocity of an observer at $p$, we have
that

\begin{description}
\item[(a)] the $p$-tangential causality of $L_{p,u}$ is spacelike.

\item[(b)] the $p$-tangential causality of $E_{p}$ is lightlike.
\end{description}
\end{proposition}

\begin{proof}

\begin{description}
\item[(a)] Given $v\in \exp _{p}^{-1}L_{p,u}$ such that $v\neq 0$,
we have
that $g\left( u,v\right) =0$. We define%
\begin{equation}
\begin{array}{crcl}
g_{u}: & T_{p}\mathcal{M} & \longrightarrow &
\mathbb{R}
\\
& v^{\prime } & \longmapsto & g_{u}\left( v^{\prime }\right)
:=g\left(
u,v^{\prime }\right) .%
\end{array}
\label{gudef}
\end{equation}%
So, $\exp _{p}^{-1}L_{p,u}=g_{u}^{-1}\left( 0\right) $.\ Therefore, given $%
w\in T_{v}\left( T_{p}\mathcal{M}\right) $, we have that $w\in
T_{v}\left( \exp _{p}^{-1}L_{p,u}\right) $ if and only if $w\left(
g_{u}\right) =0$, if and only if $g_{u}\left( \phi _{v}w\right)
=0$ (by (\ref{caniso})), if and
only if $g\left( u,\phi _{v}w\right) =0$ (by (\ref{gudef})), if and only if $%
\mathbf{g}_{p}\left( \phi _{v}^{-1}u,w\right) =0$ (by (\ref{gbold})). Then%
\begin{equation}
T_{v}\left( \exp _{p}^{-1}L_{p,u}\right) =\left( \phi
_{v}^{-1}u\right) ^{\bot }.  \label{tvexplpu}
\end{equation}%
Moreover, $\phi _{v}^{-1}u$ is timelike because
$\mathbf{g}_{p}\left( \phi
_{v}^{-1}u,\phi _{v}^{-1}u\right) =g\left( u,u\right) =-1$. Thus, (\ref%
{tvexplpu}) is a spacelike subspace and hence $\exp
_{p}^{-1}L_{p,u}$ is a spacelike submanifold of $\left(
T_{p}\mathcal{M},\mathbf{g}_{p}\right) $.

\item[(b)] Analogously,%
\begin{equation}
T_{v}\left( \exp _{p}^{-1}E_{p}\right) =\left( \phi
_{v}^{-1}v\right) ^{\bot }.  \label{tvexpep}
\end{equation}%
Since $\phi _{v}^{-1}v$ is lightlike, we have that $\exp
_{p}^{-1}E_{p}$ is a lightlike submanifold of $\left(
T_{p}\mathcal{M},\mathbf{g}_{p}\right) $.
\end{description}
\end{proof}

\section{\label{sec4}Causality of simultaneity submanifolds}

\begin{definition}
Given $p\in \mathcal{M}$ and $v\in T_{p}\mathcal{M}$, we define%
\begin{equation}
v_{q}^{\ast }:=\tau _{pq}v,  \label{adap}
\end{equation}%
where $q\in \mathcal{M}$ and $\tau _{pq}$ is the parallel
transport along
the unique geodesic segment from $p$ to $q$ (note that we are considering $%
\mathcal{M}$ as a convex normal neighborhood, and so there exists
a unique geodesic containing $p$ and $q$). It is clear that
(\ref{adap}) depends
differentiably on $q$ and thus $v^{\ast }$ is a vector field, named \textbf{%
vector field adapted to }$v$.
\end{definition}

The vector field $v^{\ast }$ adapted to $v$ has the same causal
character as $v$, since parallel transport keeps orthogonality.

\begin{definition}
Let $U$ be a timelike vector field on an open neighborhood
$\mathcal{U}$. $U$ is \textbf{synchronizable} on $\mathcal{U}$ if
and only if its orthogonal 3-distribution $U^{\bot }$ is a
foliation on $\mathcal{U}$. Then, $U^{\bot }$ is called the
\textbf{physical spaces 3-foliation of }$U$ and it is also denoted
by $S_{U}$.
\end{definition}

\begin{theorem}
\label{teocau} Given $p\in \mathcal{M}$, and $u\in
T_{p}\mathcal{M}$ the 4-velocity of an observer at $p$, we have
that

\begin{description}
\item[(a)] if $u^{\ast }$ is synchronizable on an open
neighborhood of $q\in L_{p,u}$, then $L_{p,u}$ is spacelike at
$q$.

\item[(b)] $E_{p}$ is always lightlike.
\end{description}
\end{theorem}

\begin{proof}

\begin{description}
\item[(a)] Let us call $v:=\exp _{p}^{-1}q$. By (\ref{expest}) and (\ref%
{tvexplpu}) we have
\begin{equation}
T_{q}L_{p,u}=\exp _{p\ast v}\left( T_{v}\left( \exp
_{p}^{-1}L_{p,u}\right) \right) =\exp _{p\ast v}\left( \left( \phi
_{v}^{-1}u\right) ^{\bot }\right) .  \label{tqlpu}
\end{equation}%
Let $w$ be in $\left( \phi _{v}^{-1}u\right) ^{\bot }$. Then
$g\left( u,\phi _{v}w\right) =0$, and hence $g\left( u^{\ast
},\left( \phi _{v}w\right)
^{\ast }\right) =0$ because parallel transport keeps orthogonality. So, $%
\left( \phi _{v}w\right) _{q}^{\ast }$ and $v_{q}^{\ast }$ are in
$\left( u_{q}^{\ast }\right) ^{\bot }$.

Since $u^{\ast }$ is synchronizable on an open neighborhood of
$q$, $\left( u^{\ast }\right) ^{\bot }$ is a foliation in this
open neighborhood (that is, it is involutive) and hence $\left[
v^{\ast },\left( \phi _{v}w\right) ^{\ast }\right] _{q}\in \left(
u_{q}^{\ast }\right) ^{\bot }$. Let us denote $\theta \left(
X\right) \left( Y\right) $ the Lie bracket $\left[ X,Y\right] $
where $X,Y$ are vector fields. Then $\theta \left( v^{\ast
}\right) \left( \left( \phi _{v}w\right) ^{\ast }\right) _{q}\in
\left( u_{q}^{\ast }\right)
^{\bot }$. Using induction over $n\in \mathbb{N}$%
\begin{equation}
\theta \left( v^{\ast }\right) ^{n}\left( \left( \phi _{v}w\right)
^{\ast }\right) _{q}\in \left( u_{q}^{\ast }\right) ^{\bot }.
\label{thetas}
\end{equation}%
Since
\begin{equation}
\exp _{p\ast v}w=\sum_{n=0}^{\infty }\frac{\left( -1\right)
^{n}}{\left( n+1\right) !}\theta \left( v^{\ast }\right) \left(
\left( \phi _{v}w\right) ^{\ast }\right) _{q}  \label{morcillexp}
\end{equation}
(see \cite{He}), applying (\ref{thetas}) in (\ref{morcillexp}), we have $%
\exp _{p\ast v}w\in \left( u_{q}^{\ast }\right) ^{\bot }$. So, by (\ref%
{tqlpu}), we have%
\begin{equation}
T_{q}L_{p,u}=\left( u_{q}^{\ast }\right) ^{\bot },
\label{tqlpucorta}
\end{equation}%
because they have the same dimension. Concluding, $L_{p,u}$ is
spacelike since $u_{q}^{\ast }$ is timelike.

\item[(b)] Given $q\in E_{p}$ and $v:=\exp _{p}^{-1}q$, by
(\ref{expest})
and (\ref{tvexpep}) we have%
\begin{equation}
T_{q}E_{p}=\exp _{p\ast v}\left( T_{v}\left( \exp
_{p}^{-1}E_{p}\right) \right) =\exp _{p\ast v}\left( \left( \phi
_{v}^{-1}v\right) ^{\bot }\right) .  \label{tqep}
\end{equation}%
Let $w$ be in $\left( \phi _{v}^{-1}v\right) ^{\bot }$. Then
$g\left( v,\phi _{v}w\right) =0$, and hence $g\left( v^{\ast
},\left( \phi _{v}w\right)
^{\ast }\right) =0$ because parallel transport keeps orthogonality. So, $%
\left( \phi _{v}w\right) _{q}^{\ast }$ and $v_{q}^{\ast }$ are in
$\left( v_{q}^{\ast }\right) ^{\bot }$.

Since torsion vanishes,%
\begin{equation}
\left[ v^{\ast },\left( \phi _{v}w\right) ^{\ast }\right]
_{q}=\left( \nabla _{v^{\ast }}\left( \phi _{v}w\right) ^{\ast
}\right) _{q}-\left( \nabla _{\left( \phi _{v}w\right) ^{\ast
}}v^{\ast }\right) _{q},  \label{corchete}
\end{equation}%
but
\begin{equation}
\left( \nabla _{v^{\ast }}\left( \phi _{v}w\right) ^{\ast }\right)
_{q}=0 \label{corcha}
\end{equation}
because $\left( \phi _{v}w\right) ^{\ast }$ is parallelly
transported along the geodesic that joins $p$ and $q$ (i.e., the
integral curve of $v^{\ast }$
passing through $q$). Moreover, given $X,Y,Z$ three vector fields, we have $%
Zg\left( X,Y\right) =g\left( \nabla _{Z}X,Y\right) +g\left( \nabla
_{Z}Y,X\right) $ (see \cite{He}). Taking $X,Y=v^{\ast }$ and
$Z=\left( \phi _{v}w\right) ^{\ast }$, we obtain
\begin{equation}
g\left( \nabla _{\left( \phi _{v}w\right) ^{\ast }}v^{\ast
},v^{\ast }\right) =0.  \label{corchb}
\end{equation}%
Applying (\ref{corcha}) and (\ref{corchb}) in (\ref{corchete}),
and using the previous notation for the Lie bracket, we have
$\theta \left( v^{\ast }\right) \left( \left( \phi _{v}w\right)
^{\ast }\right) _{q}\in \left( u_{q}^{\ast }\right) ^{\bot }$.
Using induction over $n\in \mathbb{N}$
\begin{equation}
\theta \left( v^{\ast }\right) ^{n}\left( \left( \phi _{v}w\right)
^{\ast }\right) _{q}\in \left( v_{q}^{\ast }\right) ^{\bot }.
\label{thetas2}
\end{equation}%
Applying (\ref{thetas2}) in (\ref{morcillexp}), we have $\exp
_{p\ast v}w\in
\left( v_{q}^{\ast }\right) ^{\bot }$. So, by (\ref{tqep}), we have%
\begin{equation}
T_{q}E_{p}=\left( v_{q}^{\ast }\right) ^{\bot },
\label{tqepcorta}
\end{equation}%
because they have the same dimension. Concluding, $E_{p}$ is
lightlike since $v_{q}^{\ast }$ is lightlike.
\end{description}
\end{proof}

Under the hypotheses of Theorem \ref{teocau}, by (\ref{tqlpucorta}) and (\ref%
{tqepcorta}), we have
\[
T_{q}L_{p,u}=\tau _{pq}u^{\bot }~~~;~~~T_{q}E_{p}=\tau
_{pq}v^{\bot },
\]
since $\left( u_{q}^{\ast }\right) ^{\bot }=\tau _{pq}u^{\bot }$
and $\left( v_{q}^{\ast }\right) ^{\bot }=\tau _{pq}v^{\bot }$.

It is important to remark that if the adapted vector field
$u^{\ast }$ is not synchronizable in any open neighborhood of $q$,
then we can not assure
that $L_{p,u}$ is spacelike at $q$, but we can always assure that the $p$%
-tangential causality of $L_{p,u}$ at $q$ is spacelike, by Proposition \ref%
{proptangcau}.

\begin{remark}
\label{remlpu} Since $L_{p,u}$ is spacelike at $p$, we can always
assure that there exists a small enough open neighborhood of $p$
in which $L_{p,u}$ is spacelike.
\end{remark}

\section{Simultaneity foliations}

Since we are supposing that the space-time $\mathcal{M}$ is a
convex normal
neighborhood, we can define Landau and horismos submanifolds at any event $p$%
. In particular, given $\beta :I\rightarrow \mathcal{M}$ an observer (where $%
I$ is an open real interval and $\beta $ is parameterized by its
proper
time), we can define the sets of Landau and horismos submanifolds%
\[
\left\{ L_{\beta \left( t\right) }\right\} _{t\in I}~~;~~\left\{
E_{\beta \left( t\right) }^{-}\right\} _{t\in I}~~;~~\left\{
E_{\beta \left( t\right) }^{+}\right\} _{t\in I}
\]%
where $L_{\beta \left( t\right) }$ denotes $L_{\beta \left( t\right) ,%
\overset{\cdot }{\beta }\left( t\right) }$. Our aim in this
section is to study these sets of Landau and horismos submanifolds
as leaves of a foliation.

\begin{figure}[tbp]
\begin{center}
\includegraphics[width=0.4\textwidth]{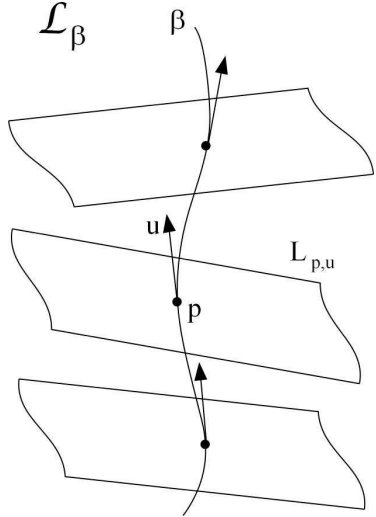}
\end{center}
\caption{Landau foliation $\mathcal{L}_{\beta }$.}
\label{figfolandau}
\end{figure}

\begin{theorem}
\label{tlandau}Let $\beta :I\rightarrow \mathcal{M}$ be an observer and $%
p\in \beta I$. There exists a convex normal neighborhood
$\mathcal{V}$ of $p$ where a foliation $\mathcal{L}_{\beta }$
(called \textbf{Landau foliation generated by }$\beta $) is
defined, and whose leaves are $\left\{ L_{\beta \left( t\right)
}\cap \mathcal{V}\right\} _{t\in I}$ (see Figure
\ref{figfolandau}).
\end{theorem}

\begin{proof}
Let us introduce some definitions that can be found in \cite{Saka96}. Let $N$%
\ be a submanifold of $\mathcal{M}$, we define the normal tangent
fiber
bundle $TN^{\bot }$\ as the fiber bundle composed by the subspaces $%
T_{p}N^{\bot }$. On an open neighborhood of the zero section
$O\left( TN^{\bot }\right) =\left\{ 0_{p}\in T_{p}N^{\bot }:p\in
N\right\} $\ we can define the normal exponential map of $N$ as
follows:
\[
\begin{array}{lrl}
\exp ^{\bot }: & TN^{\bot } & \longrightarrow \mathcal{M} \\
& v\in T_{p}N^{\bot } & \longmapsto \exp _{p}v.%
\end{array}%
\]

Considering $\beta I$\ as a submanifold of $\mathcal{M}$, we can
define the normal exponential map of $\beta I$\ on an open
neighborhood of the zero
section of $T\beta I^{\bot }$. Using an analogous reasoning given in \cite%
{Saka96}, it is proved that for all $p\in \beta I$\ there exists
an small enough open neighborhood $\mathcal{V}$ of $p$ (that we
can assume to be
convex normal neighborhood) in which the normal exponential map of $\beta I$%
\ is a diffeomorphism. So, given $q\in \mathcal{V}$, it belongs to
one and only one Landau submanifold of the family $\left\{
L_{\beta \left( t\right) }\cap \mathcal{V}\right\} _{t\in I}$.
Since these submanifolds are regular, they are the leaves of a
foliation on $\mathcal{V}$.
\end{proof}

According to Theorem \ref{tlandau} and Remark \ref{remlpu}, there
exists a small enough tubular neighborhood of $\beta I$ where the
Landau foliation generated by $\beta $ is well defined and
spacelike. So, there exists a synchronizable future-pointing
timelike unit vector field (defined on this tubular neighborhood)
orthogonal to the Landau foliation $\mathcal{L}_{\beta }$. The
integral curves of this vector field are a congruence of observers
orthogonal to $\mathcal{L}_{\beta }$.

\begin{figure}[tbp]
\begin{center}
\includegraphics[width=0.5\textwidth]{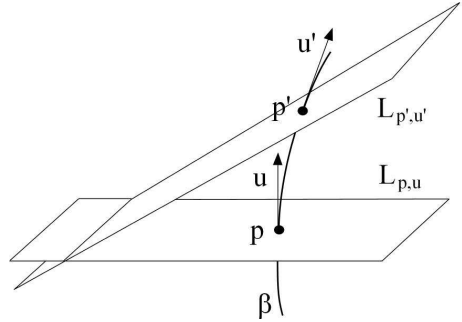}
\end{center}
\caption{Intersection of two Landau submanifolds $L_{p,u}$ and
$L_{p',u'}$ in Minkowski space-time, where $p,p'$ are events of a
not geodesic observer $\beta $ and $u,u'$ are the 4-velocities of
this observer at $p,p'$ respectively.} \label{figplanos}
\end{figure}

But, given a convex normal neighborhood, we can not assure that
the Landau foliation generated by $\beta $ is well defined on it,
because the leaves can intersect themselves. In fact, it is usual
that $L_{\beta \left(
t_{1}\right) }\cap L_{\beta \left( t_{2}\right) }\neq \emptyset $ for $%
t_{1},t_{2}\in I$, $t_{1}\neq t_{2}$; for instance, in Minkowski
space-time if $\beta $ is not geodesic (see Figure
\ref{figplanos}).

\begin{figure}[tbp]
\begin{center}
\includegraphics[width=0.25\textwidth]{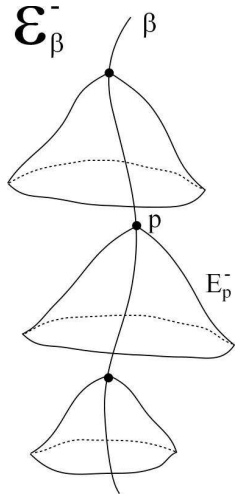}
\end{center}
\caption{Past-pointing horismos foliation $\mathcal{E}_{\beta
}^-$.} \label{figfohorismos}
\end{figure}

Unlike the Landau foliations, the horismos foliations
(past-pointing and future-pointing) are well defined on any convex
normal neighborhood, because their leaves do not intersect in any
case:

\begin{theorem}
\label{thorismos} Let $\beta :I\rightarrow \mathcal{M}$ be an
observer. There exists a foliation $\mathcal{E}_{\beta }^{-}$
(called \textbf{past-pointing horismos foliation generated by
}$\beta $) defined on $\bigcup\limits_{t\in I}E_{\beta \left(
t\right) }^{-}$ and whose leaves are $\left\{ E_{\beta \left(
t\right) }^{-}\right\} _{t\in I}$ (see Figure
\ref{figfohorismos}).

Analogous for future-pointing horismos.
\end{theorem}

\begin{proof}
It is proved in \cite{Sach77} that, in a convex normal
neighborhood, past-pointing (or future-pointing) horismos
submanifolds of different events of a given observer do not
intersect. Since they are regular sumanifolds, it is clear that
they form a foliation.
\end{proof}

According to Theorems \ref{teocau}(b) and \ref{thorismos}, the
past-pointing (and future-pointing) horismos foliation generated
by an observer is always well defined (on any convex normal
neighborhood) and lightlike.

\section{Discussion and open problems}

If we work in the framework of \textit{spacelike simultaneity}
(i.e.,
simultaneity in the local inertial proper system of the observer, see \cite%
{Oliv80}) there appear several serious mathematical problems
related with Landau submanifolds: it can not be assured that they
were spacelike at any point (see Theorem \ref{teocau}) and the
leaves of a \textquotedblleft Landau foliation\textquotedblright\
associated to a given observer can intersect themselves, even
working in a convex normal neighborhood (see Figure
\ref{figplanos}). On the other hand, all these problems disappear
if we work in the framework of \textit{lightlike} (or
\textit{observed}) \textit{simultaneity}, i.e. working with
past-pointing horismos submanifolds instead of Landau
submanifolds.

Moreover, given an observer at an event $p$ with 4-velocity $u$,
the events of its Landau submanifold $L_{p,u}$ do not affect the
observer at $p$ in any way, since both electromagnetic and
gravitational waves travel at the speed of light. On the other
hand, the events of its past-pointing horismos submanifold
$E_{p}^{-}$ are precissely the events that affect and are observed
by the observer at $p$, i.e. the events that \textit{exist} for
the observer at $p$. So, we can say \textquotedblleft \textit{what
you see is what happens\textquotedblright }. More arguments in
favor of lightlike simultaneity against spacelike simultaneity are
discussed in \cite{Bolo05}.

To conclude, we are going to present the main open problems, that
are related with Landau submanifolds:

\begin{itemize}
\item \textit{A Landau submanifold }$L_{p,u}$\textit{\ is
spacelike at any point of any convex normal neighborhood of
}$p$\textit{.}

There is not any satisfactory proof of this fact, neither a
counterexample showing that it is false.

\item \textit{Given }$\beta :I\rightarrow M$\textit{\ an observer,
we have that }$L_{\beta \left( t_{1}\right) }\cap L_{\beta \left(
t_{2}\right) }=\emptyset $\textit{\ for any }$t_{1},t_{2}\in
I$\textit{, }$t_{1}\neq t_{2} $, \textit{if and only if... ?
}(Suppossing that $\mathcal{M}$ is a convex normal neighborhood,
of course.)

In Minkowski space-time the answer is \textquotedblleft \textit{if
and only if }$\beta $\textit{\ is geodesic}\textquotedblright ,
but this property has not been yet generalized to General
Relativity. It would be useful to characterize when an observer
has focal points (see \cite{Saka96}).
\end{itemize}

\end{document}